\documentclass[conference]{IEEEtran}
\IEEEoverridecommandlockouts
\usepackage{cite}
\usepackage{bm}
\usepackage{amsmath,amssymb,amsfonts}
\usepackage{algorithm}
\usepackage{algorithmic}
\usepackage{graphicx}
\usepackage{textcomp}
\usepackage{xcolor}
\def\BibTeX{{\rm B\kern-.05em{\sc i\kern-.025em b}\kern-.08em
    T\kern-.1667em\lower.7ex\hbox{E}\kern-.125emX}}
\begin{document}

\title{Reconfigurable Intelligent Surface-Aided Spectrum Sharing Coexisting with Multiple Primary Networks\\
}
\author{\IEEEauthorblockN{Zhong Tian\IEEEauthorrefmark{1},
Zhengchuan Chen\IEEEauthorrefmark{1},
Min Wang\IEEEauthorrefmark{2},
Yunjian Jia\IEEEauthorrefmark{1},
and
Wanli Wen\IEEEauthorrefmark{1}
}
\IEEEauthorblockA{\IEEEauthorrefmark{1}
School of Microelectronics and Communication Engineering, Chongqing University, Chongqing, China\\
\IEEEauthorrefmark{2}
School of Optoelectronics Engineering, Chongqing
University of Posts and Telecommunications, Chongqing, China\\
Email: ztian@cqu.edu.cn
}
}

\maketitle

\begin{abstract}
 Considering the spectrum sharing system (SSS) coexisting with multiple primary networks, we have employed a well-designed reconfigurable intelligent surface (RIS) to control the radio environments of wireless channels and relieve the scarcity of the spectrum resource in this work. Specifically, the enhancement of the spectral efficiency of the secondary user in the considered SSS is decomposed  into two subproblems which are a second-order cone programming (SOCP) and a fractional programming of the convex quadratic form (CQFP), respectively, to optimize alternatively the beamforming vector at the secondary access point (S-AP) and the reflecting coefficients at the RIS. The SOCP subproblem is shown as a concave problem, which can be solved optimally using standard convex optimization tools. The CQFP subproblem can be solved by  a low-complexity method of gradient-based linearization with domain (GLD), providing a sub-optimal solution for fast deployment. Taking the discrete phase control at the RIS into account, a nearest point searching with penalty (NPSP) method is also developed, realizing the discretization of the phase shifts of the RIS in practice. The simulation results indicate that both GLD and NPSP can achieve an excellent performance.
\end{abstract}


\section{Introduction}
\allowdisplaybreaks
\IEEEPARstart{I}{n} 6G era, the rapid development of digital economy and the emerging applications in consumption and business stimulate the tremendous demands about the enhancement on data rate and the availability of spectrum access. Sharing spectrum among multiple networks with different priorities is a potential technology to satisfy these demands, where secondary access points (S-APs) and secondary users (SUs) are permitted to access the same spectrum and realize data transmissions without disturbing the normal communications between the primary access points (P-APs) and primary users (PUs) in primary networks (PNs) \cite{Goldsmith09CR}. Several PNs coexisting in the same spectrum band of citizen broadband radio service would worsen the data transmissions in the secondary network (SN) seriously \cite{Sohul15cbrs}. Therefore, controlling the interference from  the SN to PNs and improving the performance of the spectrum sharing system (SSS) is an urgent problem to be solved.
\par Recently, reconfigurable intelligent surface (RIS) shows the capability of  controlling and optimizing the radio environments of wireless channels between transceivers, which could improve the spectral efficiency (SE) in wireless communication systems \cite{pathlossirs2020Tang, mutigrouppan20,misostatCSI21}. Since the SSSs are interference-limited, RIS can alleviate the interference and strengthen the useful signals at the SUs by reconstructing the wireless channels in the SSSs.
The common scenarios in RIS-aided SSSs and the mathematical methods to optimize the performance of SUs were summarized and provided in \cite{zhong22VTM}, respectively.
The demo system of RIS-aided spectrum sharing was presented in \cite{indoorSRA2016TanICC} to increase the indoor network capacity significantly and relieve the overly crowded situation of spectrum resource. For the deployments of the RIS embedded in the basic SSS, the optimization for reflecting coefficients at the RIS and the power allocation of the S-AP with a single antenna  are jointly designed with successive convex approximation \cite{JPCPB2020GuanWCL}.
 The intelligent spectrum learning with the trained convolutional neural network is proposed in \cite{ISLhuang21} to reduce the interfering signals in the received signals reflected from the RIS  and maximize the signal-to-interference-plus-noise (SINR) by dynamically activating the binary status of the RIS elements.
When multiple PNs are involved in the scenarios, more complicated coupling of optimizing variables  at the S-AP and  the RIS in the SSS is introduced. Hence, the effective methods should be carefully investigated to control these variables. In particular, the discretization of the reflecting coefficients at the RIS is another concern for  practical consideration about the hardware of the RIS device.
\par In this paper, we focus on the enhancement of SE for the SSS with multiple PNs by utilizing the RIS to manipulate wireless channels, relieving the interference between PNs and the SSS with well-designed parameters.
The main contributions of this paper can be summarized as follows:
\begin{itemize}
  \item A new system model about RIS-aided SSS coexisting with multiple PNs is established, based on which an achievable rate maximization problem is formulated. With alternative optimization (AO) utilized for multiple variables, the problem  is split into two subproblems including the SOCP and the CQFP. In particular, the coupling between the RIS and the interferences is analysed clearly in the process of the problem formulation.
  \item For optimizing the beamforming vector at the S-AP, the SOCP subproblem is transformed equivalently into a concave problem and solved by standard convex optimization tools. For optimizing the reflecting coefficients of the RIS in the CQFP, we propose the method of gradient-based linearization with domain (GLD) to obtain a sub-optimal solution with a low complexity.
  \item We propose the method of nearest point searching with penalty (NPSP) to realize the discrete quantization on the phase shifts of the RIS for the practical consideration. The NPSP has little loss on the performance compared with the case with the continuous phase shifts of the RIS.
\end{itemize}
\section{System model}
\begin{figure}[!t]
  \centering
  \includegraphics[width=0.9\linewidth]{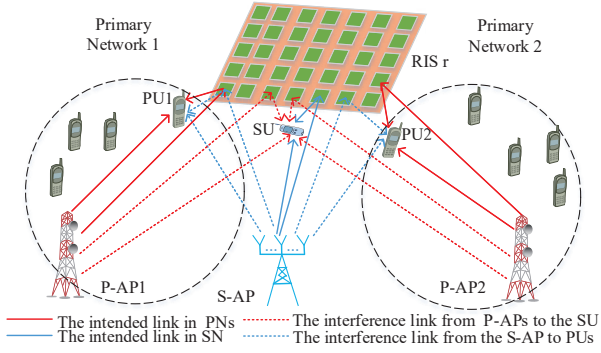}\\
  \caption{RIS-aided spectrum sharing coexisting with two primary networks.}\label{twoPUs}
  \vspace{-1.5em}
\end{figure}

In this paper, a RIS-aided SSS with $J$ PNs is considered. Fig. \ref{twoPUs} shows the scenario including two PNs. The PNs are usually controlled by the operators. Even the P-AP is configured with multiple antennas, the beamforming vector at the P-AP can not be adjusted by the SSS, whose channels to the receiver would be equivalent as an effective channel from a P-AP configured with a single antenna.  Therefore, one P-AP and several PUs (using one PU as a typical user) configured with a single antenna are assumed in each PN. Moreover, the SN includes one S-AP with $M$ antennas and one typical SU with a single antenna.
{Besides of} the direct links from the P-APs or the S-AP to the PUs or the SU, the transmit signals of the P-APs or the S-AP can also be reflected by the RIS, with the cascaded links to the PUs or the SU.

Denoting $\bm{\theta}=[\bar{\beta}_1 e^{i\bar{\phi}_1},~\dots~,~\bar{\beta}_N e^{i\bar{\phi}_N}]^\text{T}$ composed with the reflecting amplitude $\bar{\beta}_n$ and the phase shift $\bar{\phi}_n$ of $n$-th element on RIS ($n\in \mathcal{N}=\{1,\cdots,N\}$),  the reflecting coefficients of RIS are expressed as diagonal matrix $\bm{\varTheta}=\text{diag}(\bm{\theta})$.
The direct channels from a P-AP or S-AP to a PU or SU usually can be modeled as Rayleigh channel \cite{WQQ2020DFS}. Let $\text{h}_j$ and $\text{h}_{\text{p}_j,\text{b}}$ ($j\in \mathcal{J}=\{1,\cdots,J\}$) denote direct channels  from $j$-th P-AP to the corresponding PU and the SU, respectively. $\mathbf{h}_{\text{s},j}\in \mathbb{C}^{M\times 1}$ and $\mathbf{h}_\text{s}\in \mathbb{C}^{M\times 1}$ denote the direct channels from the S-AP to the PUs and the SU, respectively.
$\mathbf{h}_{\text{p}_j,\text{r}}\in \mathbb{C}^{ N\times 1}$ and $\mathbf{H}_{\text{s},\text{r}}\in \mathbb{C}^{N\times M}$ are the channels from $j$-th P-AP  and the S-AP to the RIS, respectively.
 $\mathbf{h}_{\text{r},j}\in \mathbb{C}^{ N\times 1}$ and $\mathbf{h}_\text{r,b}\in \mathbb{C}^{ N\times 1}$ refer to the channels from the RIS to $j$-th PU and the SU, respectively.
The channels between RIS and the APs or the users can be modeled as Rician fading channels \cite{WQQ2020DFS}. The received signal of the PU in $j$-th PN is expressed by
\begin{equation}\label{PUreceive}
 \begin{aligned}
  y_j&=(\mathbf{h}^\text{H}_{\text{r},j}\mathbf{\Theta }{{\mathbf{h}}_{\text{p}_j,\text{r}}}+{\text{h}_{j}})\sqrt{{{P}_{j}}}{{x}_{j}}\\
  &+(\mathbf{h}_{\text{r},j}^{\text{H}}\mathbf{\Theta }{{\mathbf{H}}_\text{s,r}}+\mathbf{h}_{\text{s},j}^{\text{H}}){{\mathbf{v}}_\text{s}}{{x}_\text{s}}+{{z}_{j}}\text{, }j\in \mathcal{J}\text{,}
 \end{aligned}
\end{equation}
where $P_j$, $x_j$, $\mathbf{v}_\text{s}$ and $z_j$ denote the transmit power of the $j$-th P-AP, the transmit symbol of the $j$-th P-AP, the beamforming vector of the S-AP and the white Gaussian noise at the $j$-th PU with $z_j\sim \mathcal{C}\mathcal{N}(0,\sigma^2_{z_j})$, respectively.
The received signal of the SU is shown as
\begin{eqnarray}
\begin{aligned}\label{SUreceive}
y_\text{b}&=(\mathbf{h}_\text{r,b}^{\text{H}}\mathbf{\Theta }{{\mathbf{H}}_\text{s,r}}+\mathbf{h}_\text{s}^{\text{H}}){{\mathbf{v}}_\text{s}}{{x}_\text{s}}\\
&+\sum\nolimits_{j=1}^{J}{(\mathbf{h}_\text{r,b}^{\text{H}}\mathbf{\Theta }{\mathbf{h}_{\text{p}_j,\text{r}}}
+{{h}_{\text{p}_j,\text{b}}})\sqrt{{{P}_{j}}}{x_{j}}}+{{z}_\text{b}}\text{,}
\end{aligned}
\end{eqnarray}
where $x_s$ and $z_b$ denote the transmit symbol of the S-AP and the white Gaussian noise at the SU with $z_b\sim \mathcal{C}\mathcal{N}(0,\sigma^2_{z_b})$, respectively.
Then, the SINR of the SU is denoted by
\begin{equation}\label{SUsinr}
  {{\gamma }_\text{b}}=\frac{|(\mathbf{h}_{\text{r},\text{b}}^{\text{H}}\mathbf{\Theta }{{\mathbf{H}}_{\text{s},\text{r}}}+\mathbf{h}_{\text{s}}^{\text{H}}){{\mathbf{v}}_\text{s}}{{|}^{2}}}{\sum_{j=1}^{J}{{{P}_{j}}|\mathbf{h}_{\text{r},\text{b}}^{\text{H}}\mathbf{\Theta }{{\mathbf{h}}_{\text{p}_j,\text{r}}}+{{h}_{\text{p}_j,\text{b}}}{{|}^{2}}}+\sigma _\text{b}^{2}}.
\end{equation}
Since the SSS is interference-limited, the noise part in ${\gamma }_\text{b}$  can be ignored. Therefore, SINR in (\ref{SUsinr}) can be simplified as SIR.
The interference  from the S-AP to the $j$-th PU is denoted as
\begin{equation}\label{INFpu}
  {\Gamma}_j=|(\mathbf{h}_{\text{r},j}^{\text{H}}\mathbf{\Theta }{\mathbf{H}_{\text{s},\text{r}}}+\mathbf{h}_{\text{s},j}^{\text{H}}){\mathbf{v}_\text{s}}|^2,\text{ }j\in \mathcal{J},
\end{equation}
where $\Gamma_j$ is restricted such that the interference from S-AP has little impact on the  normal communications in $j$-th PN. Then, the maximization of the achievable rate of the SU can be formulated as
\begin{subequations}
\begin{align}
  \text{P0:~}\max\limits_{\hat{\bm{\theta}},\mathbf{v}_\text{s}}\text{~~}&\text{log}_2\left(1+
\frac{\hat{\bm{\theta}}^\text{H}( \hat{\mathbf{H}}_\text{s}\mathbf{v}_\text{s}\mathbf{v}_\text{s}^\text{H}
\hat{\mathbf{H}}_\text{s}^\text{H})\bm{\hat{\theta}}}{\hat{\bm{\theta}}^\text{H}
(\sum_{j=1}^{J}P_{j}\bm{\tilde{\Phi}}_{j,\text{b}})\bm{\hat{\theta}}}\right)\text{~~}\label{P0ori}\\
\text{~~~~~~~~~~~s.t.~~}&|\hat{\bm{\theta}}^\text{H}\hat{\mathbf{H}}_{\text{s},j}\mathbf{v}_\text{s}|^2\leq \overline{\Gamma}_j, j\in\mathcal{J}, \text{ ~~~~~~~~~~~}\label{ctLj}\\
&|[\hat{\bm{\theta}}]_n|\leq 1,  n\in\mathcal{N},\text{~~~~~~~~~~~}\label{cttheta1N}\\
&[\hat{\bm{\theta}}]_{N+1} = 1,\text{~~~~~~~~~~~~~~~~~~}\label{ctthetaNp1}\\
&||\mathbf{v}_\text{s}||^2 \leq P_{\text{max}}, \text{~~~~~~~~~~~~~~~}\label{ctvPmax}
\end{align}
\end{subequations}
where $\hat{\bm{\theta}}^\text{H}=[\bm{\theta}^\text{H} \text{ } 1]$, $\tilde{\mathbf{h}}_{j,\text{r},\text{b}}^\text{T}=[(\text{diag}(\mathbf{h}^\text{H}_{\text{r},\text{b}})\mathbf{h}_{p_j,\text{r}})^\text{T} \text{ } h_{j,\text{b}}]$, and $\bm{\tilde{\Phi}}_{j,\text{b}}=\tilde{\mathbf{h}}_{j,\text{r},\text{b}}\tilde{\mathbf{h}}_{j,\text{r},\text{b}}^\text{H}$. $\hat{\mathbf{H}}_\text{s}$ and $\hat{\mathbf{H}}_{\text{s},j}$ are shown as
\begin{equation}
\hat{\mathbf{H}}_\text{s}=\left(
               \begin{array}{c}
                 \text{diag}(\mathbf{h}_{\text{r},\text{b}}^\text{H})\mathbf{H}_{\text{s},\text{r}}\\
                 \mathbf{h}_\text{s}^\text{H} \\
               \end{array}
             \right),\text{~}
\hat{\mathbf{H}}_{\text{s},j}=\left(
               \begin{array}{c}
                 \text{diag}(\mathbf{h}_{\text{r},j}^\text{H})\mathbf{H}_{\text{s},\text{r}}\\
                 \mathbf{h}_{\text{s},j}^\text{H} \\
               \end{array}
             \right).\nonumber
\end{equation}
\par P0 is non-concave for jointly optimizing $\mathbf{v}_\text{s}$ and $\bm{\hat{\theta}}$ because of the coupling between $\mathbf{v}_\text{s}$ and $\bm{\hat{\theta}}$ in (\ref{P0ori}) and (\ref{ctLj}). This brings a big challenge to search the optimal beamforming  vector and reflecting coefficients jointly.
In fact, even with  $\mathbf{v}_s$ fixed, the optimization on $\bm{\hat{\theta}}$ in P0 is still non-concave. Since $\hat{\mathbf{H}}_\text{s}\mathbf{v}_\text{s}\mathbf{v}_\text{s}^\text{H}
\hat{\mathbf{H}}_\text{s}^\text{H} \succeq 0$ and $\sum_{j=1}^{J}P_{j}\bm{\tilde{\Phi}}_{j,\text{b}} \succeq 0$,  the numerator and denominator in the  SIR of P0 are convex functions of $\bm{\hat{\theta}}$. Although the domain of $\bm{\hat{\theta}}$ with the constraints in P0 is a compact convex set, the fractional programming of the SIR in P0 on $\bm{\hat{\theta}}$ is convex-concave, causing the non-concave property.
\section{Alternative Optimization}
Since the non-concave property of the P0 causes the difficulty to obtain the optimal solution by optimizing beamforming vector and the reflecting coefficients jointly, we propose to optimize $\mathbf{v}_s$ and  $\bm{\hat{\theta}}$ in P0 alternatively for the sub-optimal objective value.
Although the thought of AO method is intuitional, the methods to optimize $\mathbf{v}_s$ and  $\bm{\hat{\theta}}$  need to be explored  based on the characteristics of each subproblem, respectively. Actually, the  maximization of the achievable rate in P0 is equivalent to maximizing the SIR in P0. It can be split into two subproblems for utilizing AO, which are shown as
\begin{eqnarray}
\begin{aligned}
  \text{P1:~}\text{ }\max\limits_{\mathbf{v}_\text{s}}\text{~~}&\frac{||\hat{\bm{\theta}}^\text{H} \hat{\mathbf{H}}_\text{s}\mathbf{v}_\text{s}||^2}{\hat{\bm{\theta}}^\text{H}
\left(\sum_{j=1}^{J}P_{j}\bm{\tilde{\Phi}}_{j,\text{b}}\right)\bm{\hat{\theta}}}\text{~~~}\\
  \text{~~~~~~~~~~~s.t. ~~}&\text{(\ref{ctLj}) and (\ref{ctvPmax}}),\text{~~~~~~~~}\label{P11beamv}
\end{aligned}
\end{eqnarray}
and
\begin{eqnarray}
\begin{aligned}
\text{P2:~}\max\limits_{\hat{\bm{\theta}}}\text{~~}&\frac{\hat{\bm{\theta}}^H( \hat{\mathbf{H}}_\text{s}\mathbf{v}_\text{s}\mathbf{v}_\text{s}^\text{H}
\hat{\mathbf{H}}_\text{s}^\text{H})\bm{\hat{\theta}}}{\hat{\bm{\theta}}^\text{H}
\left(\sum_{j=1}^{J}P_{j}\bm{\tilde{\Phi}}_{j,\text{b}}\right)\bm{\hat{\theta}}}\\
  \text{s.t.~~}&\text{(\ref{ctLj}), (\ref{cttheta1N}) and (\ref{ctthetaNp1})}.\text{~}\label{P12theta}
\end{aligned}
\end{eqnarray}
In the subsequent subsections, we investigate the solutions of P1 and P2, respectively.
\subsection{ The Optimal Beamforming Vector at the S-AP}
With the variable $\hat{\bm{\theta}}$ fixed in  P1, the SIR in P0 is optimized with the beamforming vector $\mathbf{v}_\text{s}$ of S-AP in P1. Since the value of objective function in P1 remains the same with $\mathbf{v}_s$ replaced by $e^{i\phi}\mathbf{v}_\text{s}$ ($\forall \phi \in[0,2\pi)$), P1 is equivalent to P3 shown as
\begin{eqnarray}
\begin{aligned}
\text{P3:~ }\max\limits_{\mathbf{v}_\text{s}}\text{~~}&\frac{\text{Re}\left(\hat{\bm{\theta}}^\text{H} \hat{\mathbf{H}}_\text{s}\mathbf{v}_\text{s}\right)}{\hat{\bm{\theta}}^\text{H}\left(\sum_{j=1}^{J}
P_{j}\bm{\tilde{\Phi}}_{j,\text{b}}\right)\bm{\hat{\theta}}}\\
  \text{~~~~~s.t.~~}&\text{(\ref{ctLj}), (\ref{ctvPmax})  and }
\text{Im}(\hat{\bm{\theta}}^\text{H} \hat{\mathbf{H}}_\text{s}\mathbf{v}_\text{s})=0,\text{~}\label{P21beamv}
\end{aligned}
\end{eqnarray}
where $\text{Re}(\cdot)$ and $\text{Im}(\cdot) $ denote the real part and the imaginary part of a scalar, a vector or a matrix, respectively.
\par For the constraints of $\mathbf{v}_\text{s}$ in P3, the feasible domain of $\mathbf{v}_\text{s}$ is a convex set. Besides, $\mathbf{v}_\text{s}$ in (\ref{P21beamv}) is a concave function with known $\hat{\bm{\theta}}$. This implies that P3 can be verified as the maximization of a concave function on $\mathbf{v}_\text{s}$ with the form of SOCP \cite{IRS2020LiangTcom}, which can be solved with the common algorithms of convex optimization (such as interior point method). Therefore, we can obtain the optimal solution from optimizing the beamforming vector at the S-AP in the AO with the reflecting coefficients fixed.
\subsection{The Optimization of the Reflecting Coefficients }
In the design of the algorithm for CQFP, the computation complexity is important for the application in the practical system. To solve P2 in low complexity, the sub-optimal GLD is proposed to optimize the reflecting coefficients at the RIS.  The general idea of GLD can be described as follows. First, P2 would be transformed as the minimization of a concave function. Then, linearization with domain would  be utilized to obtain the local-optimal solution for the minimization of a concave problem, which is a sub-optimal solution for P2.
\begin{subequations}
\begin{align}
   \text{P4:~}\min\limits_{\hat{\bm{\theta}}, t}\text{~~}&-\displaystyle\frac{||(\hat{\mathbf{H}}_\text{s}
\mathbf{v}_\text{s})^\text{H}\bm{\hat{\theta}}||^2}{t}\label{P4here}\\
\text{s.t.~~}&{\hat{\bm{\theta}}^\text{H}({\sum}_{j=1}^{J}P_{j}\bm{\tilde{\Phi}}_{j,\text{b}})
\bm{\hat{\theta}}}\leq t, \text{}\label{tgreqxBx}\\
&\text{(\ref{ctLj}), (\ref{cttheta1N}), and (\ref{ctthetaNp1})}. \nonumber
\end{align}
\end{subequations}
The problem P2 can be transformed into P4 equivalently, which is the minimization of a concave problem.
The equivalence between P2 and P4 can be proven as follows.
\begin{IEEEproof}
If $\hat{\bm{\theta}}_0$ is the optimal solution for P2, define $t_0={\hat{\bm{\theta}}_0^\text{H}(\sum_{j=1}^{J}P_{j}
\bm{\tilde{\Phi}}_{j,\text{b}})\bm{\hat{\theta}}_0}$. Then, ($\hat{\bm{\theta}}_0,t_0$) is a feasible solution for P4. Assuming ($\hat{\bm{\theta}}_0,t_0$) is not the optimal solution for P4, the optimal solution ($\hat{\bm{\theta}}_1,t_1$) of P4 exists and ${||(\hat{\mathbf{H}}_\text{s}\mathbf{v}_\text{s})^\text{H}\bm{\hat{\theta}}_1||^2}/{t_1}>
{||(\hat{\mathbf{H}}_\text{s}\mathbf{v}_\text{s})^\text{H}\bm{\hat{\theta}}_0||^2}/{t_0}$.
Therefore,
\begin{equation}
\frac{||(\hat{\mathbf{H}}_\text{s}\mathbf{v}_\text{s})^\text{H}\bm{\hat{\theta}}_1||^2}
{\hat{\bm{\theta}}_1^\text{H}
(\sum\limits_{j=1}^{J}P_{j}\bm{\tilde{\Phi}}_{j,\text{b}})\bm{\hat{\theta}}_1}\geq
\frac{||(\hat{\mathbf{H}}_\text{s}\mathbf{v}_\text{s})^\text{H}\bm{\hat{\theta}}_1||^2}{t_1}> \frac{||(\hat{\mathbf{H}}_\text{s}\mathbf{v}_\text{s})^\text{H}\bm{\hat{\theta}}_0||^2}
{\hat{\bm{\theta}}_0^\text{H}
(\sum\limits_{j=1}^{J}P_{j}\bm{\tilde{\Phi}}_{j,\text{b}})\bm{\hat{\theta}}_0}\nonumber
\end{equation}
exists, which means that $\hat{\bm{\theta}}_1$ is {a} feasible solution for P2 with higher value for objective function of P2, compared with the optimal solution $\hat{\bm{\theta}}_0$ for P2. This results in the contradiction with the definition of $\hat{\bm{\theta}}_0$ for P2. Therefore, ($\hat{\bm{\theta}}_0,t_0$) is also an optimal solution for P4.

If ($\hat{\bm{\theta}}_2,t_2$) is the optimal solution for P4, $t_2={\hat{\bm{\theta}}_2^\text{H}(\sum_{j=1}^{J}P_{j}\bm{\tilde{\Phi}}_{j,\text{b}})\bm{\hat{\theta}}_2}$ can be  proven. The explanation is illustrated as follows. If $t_2\neq {\hat{\bm{\theta}}_2^\text{H}(\sum_{j=1}^{J}P_{j}\bm{\tilde{\Phi}}_{j,\text{b}})\bm{\hat{\theta}}_2}$,  $t_2>{\hat{\bm{\theta}}_2^\text{H}(\sum_{j=1}^{J}P_{j}\bm{\tilde{\Phi}}_{j,\text{b}})\bm{\hat{\theta}}_2}$ is obtained with the constraints  (\ref{tgreqxBx}). Then, a smaller value for objective function of P4 with solution ($\hat{\bm{\theta}}_2,\hat{t}_2$) is obtained, where $\hat{t}_2={\hat{\bm{\theta}}_2^\text{H}(\sum_{j=1}^{J}P_{j}\bm{\tilde{\Phi}}_{j,\text{b}})\bm{\hat{\theta}}_2}$.
This also results in the contradiction with the definition of ($\hat{\bm{\theta}}_2,t_2$) for P4. With $t_2={\hat{\bm{\theta}}_2^\text{H}(\sum_{j=1}^{J}P_{j}\bm{\tilde{\Phi}}_{j,\text{b}})\bm{\hat{\theta}}_2}$, $\hat{\bm{\theta}}_2$  is the optimal solution for P4$'$ shown as
\begin{align}
  \text{P4}'\text{:~}\min\limits_{\hat{\bm{\theta}}, t}\text{~~}&-
  \frac{||(\hat{\mathbf{H}}_\text{s}\mathbf{v}_\text{s})^\text{H}\bm{\hat{\theta}}||^2}{t} \nonumber\\
\text{s.t.~~~}&{\hat{\bm{\theta}}^\text{H}\left(\sum\nolimits_{j=1}^{J}P_{j}\bm{\tilde{\Phi}}_{j,\text{b}}\right)
\bm{\hat{\theta}}}= t\text{,} \nonumber\\
&\text{(\ref{ctLj}), (\ref{cttheta1N}), and (\ref{ctthetaNp1})}. \nonumber\label{P41}
\end{align}
The equivalence between P2 and P4$'$ is obvious. With all the derivations and illustrations shown, P2 is equivalent to P4.
\end{IEEEproof}
Since $f(\bm{x},t)=||\bm{x}||^2/t$ is convex over ($\bm{x}, t\in \Re_+$), $q(\bm{\hat{\theta}},t)\triangleq -||(\hat{\mathbf{H}}_\text{s}\mathbf{v}_\text{s})^\text{H}\bm{\hat{\theta}}||^2/t$ is concave over ($\bm{\hat{\theta}}, t\in \Re_+$). As P4 is the minimization over the concave problem, the common methods in convex optimization to solve the minimization over a convex function for the optimal solution could not be utilized. However, the approximation of the concave objective function in P4 with linearization can be considered to obtain a sub-optimal solution for P4. Linear approximation is the variation of Taylor expansion by {cutting off the Lagrange form of the remainder or the Peano form of the remainder}, which is in second order. Therefore, the value of the linear approximation for the objective function in P4 is higher than the optimal value in P4 for the convexity of the function $||\bm{x}||^2/t$, which can explain the sub-optimality of GLD.
\par With $\bm{\varpi}^\text{T}=[\bm{\hat{\theta}}^\text{T} ~ t]$ defined, linearization is utilized to make $q(\bm{\varpi})$ convexified on $\bm{\varpi}_k$ as
\begin{equation}\label{qlin}
  \hat{q}(\bm{\varpi};\bm{\varpi}_k)\triangleq q(\bm{\varpi}_k)+\nabla q(\bm{\varpi}_k)^\text{T}(\bm{\varpi}-\bm{\varpi}_k),
\end{equation}
where $\hat{q}(\bm{\varpi};\bm{\varpi}_k)$ is an affine function with convexity and $k$ is the count number of the iteration in the GLD. The domain of $\bm{\varpi}$ is $\mathcal{D}$ determined by the constraints (\ref{ctLj}), (\ref{cttheta1N}), (\ref{ctthetaNp1}) and (\ref{tgreqxBx}), and the convexity of the domain can be easily validated.
\par To keep any feasible point for the problem (\ref{qlin}) in $\mathcal{D}$, the penalty item should be added in (\ref{qlin}) for the points out of $\mathcal{D}$. The linearization for P4 with domain is shown as
\begin{eqnarray}\label{qlinDomain}
  &\tilde{q}(\bm{\varpi};\bm{z})\triangleq q(\bm{z})+\nabla q(\bm{z})^\text{T}(\bm{\varpi}-\bm{z})+\mathcal{I}(\bm{\varpi}),\\
  &\mathcal{I}(\bm{\varpi})\triangleq \left\{
  \begin{array}{lr}
  0& \bm{\varpi}\in \mathcal{D}, \\
  \infty& \bm{\varpi}\notin \mathcal{D}.
  \end{array}
  \right.\text{~~~~~~~~~~~~~~~~}\nonumber
\end{eqnarray}
Since $\mathcal{I}(\bm{\varpi})$ is a convex function  and (\ref{qlin}) is an affine function,  $\tilde{q}(\bm{\varpi};\bm{z})$ in (\ref{qlinDomain}) is convex.
The gradient of $q(\bm{\varpi})$ can be obtained as
\begin{equation}\label{gradientq}
  \nabla q(\bm{\varpi})^\text{T}=[-\hat{\bm{\theta}}^\text{H}(\hat{\mathbf{H}}_\text{s}\mathbf{v}_\text{s}\mathbf{v}_\text{s}^\text{H}\hat{\mathbf{H}}_\text{s}^\text{H})/t \text{~ }||(\hat{\mathbf{H}}_\text{s}\mathbf{v}_\text{s})^\text{H}\bm{\hat{\theta}}||^2/t^2].
\end{equation}
Since the optimization of (\ref{qlin}) includes the process of iteration for $\bm{\varpi}_k$, $\bm{\varpi}_k$ may be on the boundary of the $\mathcal{D}$. Therefore, the convex function $\hat{q}(\bm{\varpi};\bm{\varpi}_k)$ may not be differentiable at $\bm{\varpi}_k$. If $\bar{\bm{\varpi}}_k$ in $q(\bar{\bm{\varpi}}_k)$ is on the boundary of $\mathcal{D}$, the damped process as $\bm{\varpi}_k=\varepsilon \bar{\bm{\varpi}}_k+(1-\varepsilon)\bm{\varpi}_{k-1}$ is executed with $\varepsilon \in (0\text{, } 1)$. If $\bm{\varpi}_{0}$ is in the interior of $\mathcal{D}$, $\bm{\varpi}_k$ will be guaranteed in the interior of $\mathcal{D}$.
\par In the initial phase, the GLD attempts to obtain a feasible solution for P4 randomly. With this feasible point in the domain of P4, the objective function in P4 would be  approximated by a line segment with the linearization operation. The optimal solution for the line segment in the feasible domain of P4 is solved as the updated feasible point to start next iteration.  Therefore, the iteration of the $\bm{\varpi}_k$ would converge to a local minimum point of the objective function in P4, which is a sub-optimal solution of the reflecting coefficients at the RIS. In summary, the GLD algorithm is shown as \textbf{Algorithm \ref{alggld}}, whose computation complexity is proportional to $\mathcal{O}(\overline{K}N)$.
\begin{algorithm}
\caption{The Gradient-based Linearization with Domain algorithm for P4}
\begin{algorithmic}[1]\label{alggld}
\STATE Initial: given $\bm{\varpi}_0$ in interior of $\mathcal{D}$ randomly, $\tau_0=q(\bm{\varpi}_0)$,  $\tau_{-1}=-\infty$, $k\leftarrow 0$, set $\overline{K}$ with a large positive integer.
\WHILE {$\tau_k>\tau_{k-1}$\& $k<\overline{K}$}
\STATE Linearization with domain: $\tilde{q}(\bm{\varpi};\bm{\varpi}_k)=q(\bm{\varpi}_k)+\nabla q(\bm{\varpi}_k)^T(\bm{\varpi}-\bm{\varpi}_k)+\mathcal{I}(\bm{\varpi})$.
\STATE Solve the convex subproblem of minimizing $\tilde{q}(\bm{\varpi};\bm{\varpi}_k)$ with interior point method and obtain the solution $\bar{\bm{\varpi}}_{k+1}$  for $\tau_{k}=\min_{\bm{\varpi}}\tilde{q}(\bm{\varpi};\bm{\varpi}_k) $with the constraints (\ref{ctLj}), (\ref{cttheta1N}), (\ref{ctthetaNp1}) and (\ref{tgreqxBx}).
\IF {$\bar{\bm{\varpi}}_{k+1}$ is in the interior of $\mathcal{D}$}
\STATE $\bm{\varpi}_{k+1}\leftarrow \bar{\bm{\varpi}}_{k+1}$
\ELSE
\STATE $\bm{\varpi}_{k+1}\leftarrow \varepsilon \bar{\bm{\varpi}}_k+(1-\varepsilon)\bm{\varpi}_k$, $\varepsilon \in (0\text{, } 1)$.
\ENDIF
\STATE $k\leftarrow k+1$
\ENDWHILE
\end{algorithmic}
\end{algorithm}
\section{The Practical Consideration for the RIS}
Although the phase shifts of reflecting coefficients in P0 are continuous variables ranging in $[0\text{, }2\pi)$, the phase shifts of the RIS in practical system are selected from discrete values for the characteristics of the RIS devices. Usually, the phase shifts can be quantified by 1 bit or 2 bits \cite{WQQ2020DFS}. If the phase shifts of $\hat{\bm{\theta}}$ in P0 are discrete variables, P0 will become a mixed integer non-linear problem (MINLP), which is more complicated to be solved.  The discrete phase shifts of $\hat{\bm{\theta}}$ can be relaxed to continuous variables to obtain the solution, which is corresponding to P0. Then, the continuous solution can be quantified to discrete phase shifts. However, new solution should be feasible for the constraints (\ref{ctLj}) and decrease the deterioration of SE performance. Assuming that $\bar{\bm{\theta}}$ is a continuous solution for P0 and there are $L$ levels ($\{\phi_1,\cdots,\phi_L\}$) for the quantization of phase shifts, the quantization problem about the phase shifts of the reflecting coefficients can be illustrated as
\begin{subequations}
\begin{align}
\text{ P5: }\min\limits_{\bm{\theta}_\text{d}}\text{~~}&||\bar{\bm{\theta}}-\bm{\theta}_\text{d}||^2 \label{P5here}\\
\text{s.t.~~} &\angle([\bm{\theta}_\text{d}]_n)\in  \{ \phi_1,\cdots,\phi_L \}, n\in \mathcal{N},\text{~~~~}\label{thetadrange}\\
&|\bm{\theta}_\text{d}^\text{H}\hat{\mathbf{H}}_{\text{s},j}\mathbf{v}_\text{s}|^2\leq \overline{\Gamma}_j, j\in\mathcal{J}. \label{dctLj}
\end{align}
\end{subequations}
For a small number of reflecting coefficients, the exhaustive searching method can be utilized to obtain the discrete solution $\bm{\theta}_\text{d}$, which should conform to the constraints of (\ref{ctLj}) and optimize the objective function of P0.
\par In this paper, the NPSP is proposed to obtain the sub-optimal solution about the discrete phase shifts. The NPSP method would focus on searching the nearest point in the candidate set of the discrete phase shifts, where the objective function would be added with the penalty for invalidation of constraints. Let us introduce an auxiliary vector $\textbf{b}$ for $\bm{\theta}_\text{d}$. Then, the problem about NPSP for solving discrete phase shifts at the RIS is formulated as
\begin{equation}
\begin{aligned}
   \text{ P6: }\min\limits_{\bm{\theta}_\text{d}\text{, }\textbf{b}}\text{~~}&||\bar{\bm{\theta}}-\textbf{b}||^2+\frac{\mu}{2}||\textbf{b}-\bm{\theta}_\text{d}||^2
   \\
\text{s.t.~~} &\textbf{b}=\bm{\theta}_\text{d},\text{~} \text{(\ref{thetadrange}), and (\ref{dctLj})}, \label{P6here}
\end{aligned}
\end{equation}
where $\mu>0$ is the penalty parameters. Given the Lagrange variables $\textbf{w}=[w_1,\cdots,w_J]^\text{T}\succeq 0$  and $\bm{\lambda}=\bm{\lambda}_\text{R}+\text{i}\bm{\lambda}_\text{I}$ ($\bm{\lambda}_\text{R}=[\lambda_{\text{R},1},\cdots,\lambda_{\text{R},N}]^\text{T}$ and $\bm{\lambda}_\text{I}=[\lambda_{\text{I},1},\cdots,\lambda_{\text{I},N}]^\text{T}$) for $\text{Re}\{\textbf{b}-\bm{\theta}_\text{d}\}=0$ and $\text{Im}\{\textbf{b}-\bm{\theta}_\text{d}\}=0$, the Lagrangian function for P6 is
\begin{eqnarray}\label{langFun}
\begin{aligned}
  \mathcal{G}(\textbf{b},\bm{\theta}_\text{d},\bm{\lambda},\textbf{w})=&
  (\textbf{b}-\bar{\bm{\theta}})^\text{H}(\textbf{b}-\bar{\bm{\theta}})+
  \frac{\mu}{2}(\textbf{b}-\bm{\theta}_\text{d})^\text{H}(\textbf{b}-\bm{\theta}_\text{d})\\
  +&\sum\nolimits_{n=1}^N\mathcal{I}_n([\bm{\theta}_\text{d}]_n)+
  \text{Re}\{\bm{\lambda}^\text{H}(\textbf{b}-\bm{\theta}_\text{d})\}\\
  +&\sum\nolimits_{j=1}^Jw_j(|\bm{\theta}_\text{d}^\text{H}\hat{\mathbf{H}}_{\text{s},j}\mathbf{v}_\text{s}|^2-\overline{\Gamma}_j),
\end{aligned}
\end{eqnarray}
where  $\mathcal{I}_n (\cdot)$ is the indicator function ($\mathcal{I}_n(x)=1$ if $x\in \{ |[\bar{\bm{\theta}}]_n|e^{\text{i}\phi_1},\cdots,|[\bar{\bm{\theta}}]_n|e^{\text{i}\phi_L} \}$, otherwise $+\infty$). Then, the dual problem can be defined as
\begin{equation}\label{dualprob}
 \mathcal{L}(\bm{\lambda},\textbf{w})\triangleq \inf \limits_{\textbf{b},\bm{\theta}_\text{d}}{\mathcal{G}(\textbf{b},\bm{\theta}_\text{d},\bm{\lambda},\textbf{w})}
\end{equation}
To solve P6 with the  NPSP, the iterations about the parameters are shown as
\begin{subequations}
\begin{align}
 &\bm{\theta}^{(k+1)}_\text{d}=\text{arg}\min\limits_{\bm{\theta}_\text{d}}\mathcal{G}(\textbf{b}^{(k)},\bm{\theta}_\text{d},\bm{\lambda}^{(k)},\textbf{w}^{(k)}),\label{thetaditer}\\
 &\textbf{b}^{(k+1)}=\text{arg}\min\limits_{\textbf{b}}\mathcal{G}(\textbf{b},\bm{\theta}_\text{d}^{(k+1)},\bm{\lambda}^{(k)},\textbf{w}^{(k)}),\label{qiter}\\
 &\bm{\lambda}^{(k+1)}=\bm{\lambda}^{(k)}+\mu(\textbf{b}^{(k+1)}-\bm{\theta}_\text{d}^{(k+1)}),\label{lambdaiter}\\
 &w^{(k+1)}_j=w^{(k)}_j+\mu(|\bm{\theta}^{(k+1)}_\text{d}\hat{\mathbf{H}}_{\text{s},j}\mathbf{v}_\text{s}|^2-\overline{\Gamma}_j)^+\text{, }j\in \mathcal{J}.\label{witer}
\end{align}
\end{subequations}																									
Since $(\mathbf{H}_{\text{s},j}\mathbf{v}_\text{s})^\text{H}(\mathbf{H}_{\text{s},j}\mathbf{v}_\text{s})$ is a real number,  the angle calculation is shown as
\begin{equation}\label{findthetakp1}
  \angle\left(\frac{\textbf{b}^{(k)}+\bm{\lambda}^{(k)}/\mu}
  {1+\sum_{j=1}^Jw_j(\mathbf{H}_{\text{s},j}\mathbf{v}_\text{s})^\text{H}(\mathbf{H}_{\text{s},j}\mathbf{v}_\text{s})}\right)
  =\angle\left(\textbf{b}^{(k)}+\frac{\bm{\lambda}^{(k)}}{\mu}\right).\nonumber
\end{equation}
Therefore, $\bm{\theta}^{(k+1)}_\text{d}$ in (\ref{thetaditer}) can be solved by
\begin{eqnarray}\label{itertheta}
  &\angle(\bm{\theta}^{(k+1)}_\text{d})\triangleq \text{arg}\min\limits_{\bm{\varphi}}||\bm{\varphi}-\angle\left(\textbf{b}^{(k)}+\bm{\lambda}^{(k)}/\mu\right)||,\\
  &[\bm{\varphi}]_n\in\{\phi_1,\cdots,\phi_L\}\text{, } |[\bm{\theta}^{(k+1)}_\text{d}]_n|\triangleq [\bar{\bm{\theta}}]_n\text{, } n\in\mathcal{N}.\nonumber
\end{eqnarray}
Besides, $\textbf{b}^{(k+1)}$ in (\ref{qiter}) can be solved by
\begin{eqnarray}\label{qiterreal}
\begin{aligned}
  \textbf{b}^{(k+1)}=\textbf{Y}^{-1}\left(2\bar{\bm{\theta}}+\mu\bm{\theta}_\text{d}^{(k+1)}+\bm{\lambda}^{(k+1)}\right),\\
\end{aligned}
\end{eqnarray}
where $\textbf{Y}=(2+\mu)\textbf{I}_N+2\sum_{j=1}^Jw_j^{(k)}
  (\mathbf{H}_{\text{s},j}\mathbf{v}_\text{s})(\mathbf{H}_{\text{s},j}\mathbf{v}_\text{s})^\text{H}$.
Then, the proposed NPSP algorithm to solve P6 problem can be shown as \textbf{Algorithm \ref{npsp}}, whose computation complexity is  proportional to $\mathcal{O}(N_\text{itr}(N(L+1)+JM+N^3))$.
\begin{algorithm}
\caption{The Nearest Point Searching with Penalty algorithm for P6}
\begin{algorithmic}[1]\label{npsp}
\STATE Initial: given $\bm{\lambda}^{(0)}$, $\textbf{b}^{(0)}$ and $\textbf{w}^{(0)}$  randomly, $F_\text{obj}^{(0)}\leftarrow +\infty$,  $k\leftarrow 0$. Set the terminating threshold $\varsigma$ for objective function in P5 and $N_\text{itr}$ for maximum number of iteration.
\WHILE {$F_\text{obj}^{(k)}>\varsigma$\text{~}\&\text{~}$k<N_\text{itr}$}
\STATE Update $\bm{\theta}_\text{d}^{(k+1)}$, $\textbf{b}^{(k+1)}$ with (\ref{itertheta})
and (\ref{qiterreal})
\IF {$\bm{\theta}_\text{d}^{(k+1)}$  meets the constraints in (\ref{dctLj})}
\IF{$||\bar{\bm{\theta}}-\bm{\theta}_\text{d}^{(k+1)}||^2<F_\text{obj}^{(k)}$}
\STATE $\bm{\theta}_\text{o}\leftarrow \bm{\theta}_\text{d}^{(k+1)}$
\ENDIF
\STATE $F_\text{obj}^{(k+1)}\leftarrow \min\{||\bar{\bm{\theta}}-\bm{\theta}_\text{d}^{(k+1)}||^2,F_\text{obj}^{(k)}\}$

\ENDIF
\STATE Update $\bm{\lambda}^{(k+1)}$, $\textbf{w}^{(k+1)}$ with (\ref{lambdaiter}) and (\ref{witer})
\STATE $k\leftarrow k+1$
\ENDWHILE
\IF {$F_\text{obj}^{(k)}=+\infty$\text{~}\&\text{~}$k=N_\text{itr}$}
\STATE The feasible solution for $\bm{\theta}_\text{d}$ is not searched.
\ELSE
\STATE Output $\bm{\theta}_\text{o}$
\ENDIF
\end{algorithmic}
\end{algorithm}
\par Since $\bm{\theta}_\text{d}$ is the discretization of the $\bar{\bm{\theta}}$, the constraints of (\ref{dctLj}) may become invalid. Therefore, the feasible $\bm{\theta}_\text{d}$ may not exist for some CSI $\mathbf{H}_{\text{s},j}$ and the beamforming vector $\mathbf{v}_\text{s}$, especially with a small number of elements at the RIS.

\section{Simulation Results and Discussions}
\begin{figure}[htpb]
  \centering
  \includegraphics[width=0.9\linewidth]{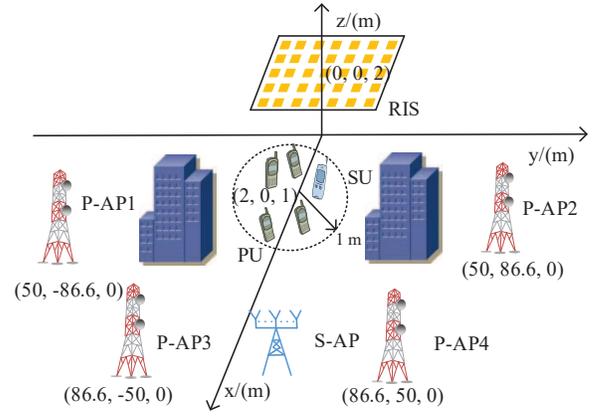}\\
  \caption{Simulated scenarios: a RIS-aided SSS coexisting with multiple PNs. }\label{OneSU4PUs}
  \vspace{-1.em}
\end{figure}
In this section, simulation results are shown to evaluate the performance of RIS-aided spectrum sharing with multiple PNs.
The general simulation setup is shown in Fig. \ref{OneSU4PUs}.
The path losses of direct link and the cascaded link over the RIS are 106 dB and 123 dB, respectively, which are calculated by following the parameters given in Table \ref{tableax} \cite{weightsumr2020guo}. In particular, due to the double fading effect \cite{pathlossirs2020Tang}, the path loss of the cascaded link over RIS is larger than that of the direct link.

\begin{table}[ht]
\centering
\caption{The Parameters of the Simulation}
\begin{tabular}{cc}
  \hline
  Parameters & Values \\
  \hline
  Path loss for RIS-assisted channel & 35.6+22.0 lg d \\
  Path loss for direct channel & 32.6+36.7 lg d \\
  The maximum transmit power ($P_\text{max}$) &-2 dBm $\sim$ 14 dBm\\
  The interference threshold ($\overline{\Gamma}_j$) &-115 dBm, $j\in \mathcal{J}$\\
  The power of a P-AP ($P_j$) &10 dBm, $j\in \mathcal{J}$\\
  \hline
\end{tabular}
\label{tableax}
\end{table}
\par
\begin{figure}[!t]
  \centering
  \includegraphics[width=0.75\linewidth]{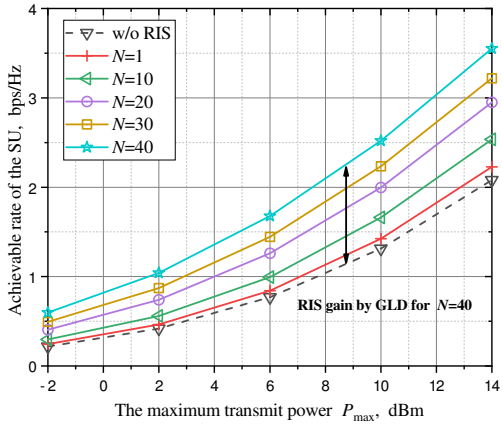}\\
  \caption{The performance of the SU realized by the GLD method as $N$ varies.}\label{fullsld}
  \vspace{-1em}
\end{figure}

  Fig. \ref{fullsld} shows the SE performance of the proposed GLD with the varying parameter $N$ of the RIS as $P_\text{max}$ of the S-AP increases from $-2\text{ dBm} $ to $14 \text{ dBm}$. Compared with the scenario without RIS, RIS-aided SSS  has obvious gain on the SE performance, even with $N=1$. For the objective function of P2, the increasing on the parameter $N$ of the RIS not only provide more flexibility on the optimization of  the numerator in P2, may also amplify or diminish the denominator in P2. As Fig. \ref{fullsld} shows, our proposed GLD  improves the achievable rate of the SU, which indicates that the increasing on $N$ can enhance the SE performance of the system.  When $P_\text{max}$ ranges in the interval of high power , the proposed GLD method can achieve better SE of the system as $N$ becomes larger.
\par
\begin{figure}[!t]
  \centering
  \includegraphics[width=0.75\linewidth]{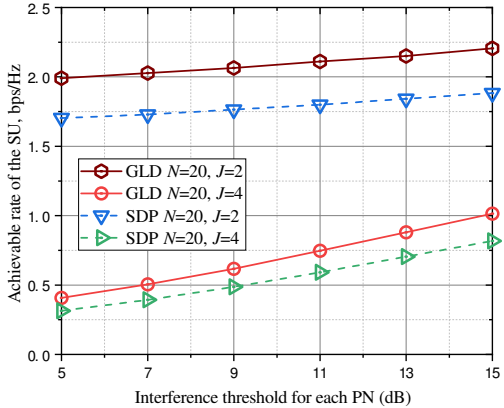}\\
  \caption{The performance of the SU obtained by the GLD method and the benchmark SDP method versus the interference threshold of each PN with $P_\text{max}=10 \text{ dBm}$.}\label{diffusers}
\vspace{-1.5em}
\end{figure}

 Fig. \ref{diffusers} shows the SE performance of the proposed GLD and the benchmark semidefinite programming (SDP) method as the interference threshold of each PN $\overline{\Gamma}_j$ increases. With more PNs involved in the scenarios of spectrum sharing, the SE performance of the system decreases sharply for more interference introduced in the denominator of $\gamma_b$ and more constraints added in (\ref{ctLj}), which is shown in Fig. \ref{diffusers}. Besides, the enhancement on the SE performance of the SU becomes more obvious for the scenarios with a larger number of PNs as the interference threshold goes up.
\par
\begin{figure}[htpb]
  \centering
  \includegraphics[width=0.75\linewidth]{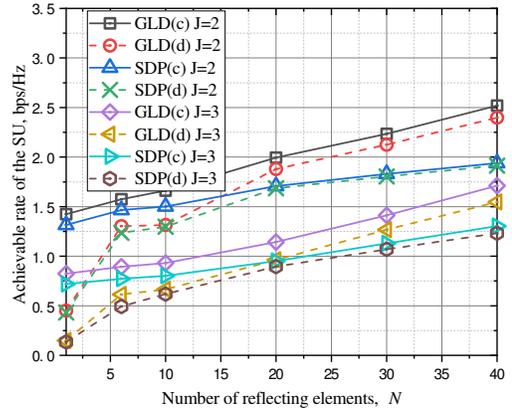}\\
  \caption{The phase discretization of $\bm{\hat{\theta}}$ by NPSP with $P_\text{max}=10 \text{ dBm}$ and 2 bits used to quantify the phase shifts where (c) and (d) mean continuous value and discrete value for $\bm{\hat{\theta}}$ with $\overline{\Gamma}_i=5\text{~dB}$, respectively.}\label{dthetarate}
  \vspace{-1.5em}
\end{figure}

 For the practical consideration about the reflecting coefficients, the proposed NPSP is utilized to quantize the phase shifts of the reflecting coefficients $\bm{\hat{\theta}}$ at the RIS for the proposed GLD  and the benchmark SDP method. Fig. \ref{dthetarate} shows the performance of GLD and SDP with the embedded NPSP for the phase discretization of $\bm{\hat{\theta}}$. The discretization of $\bm{\hat{\theta}}$ decreases the achievable rate of the SU slightly for the proposed GLD  and the benchmark SDP method with $J=2$ or $3$. Besides, the achievable rate of SU about the proposed GLD  with NPSP embedded for the phase discretization of $\bm{\hat{\theta}}$ is still better than that of benchmark SDP method in the same case. It is worth noting that the achievable rate of the SU with the phase discretization of $\bm{\hat{\theta}}$ encounters the sharp decreasing with a small $N$, compared with that of continuous value of $\bm{\hat{\theta}}$. When $N$ is small, the phase discretization of $\bm{\hat{\theta}}$ may result in the invalidation about the constraints in (\ref{dctLj}).
\section{Conclusion}
In this paper, the problem about maximizing the achievable rate of the SU in RIS-aided SSS with multiple PNs was formulated, which was split into a SOCP subproblem and a CQFP subproblem, respectively. The SOCP subproblem has been proven to be concave, which can obtain the optimal solution with the common algorithms of convex optimization. To decrease the computation complexity, we developed the sub-optimal GLD to solve the CQFP subproblem. Although more PNs would decrease the SE performance of the SU due to more interference introduced in the system, our proposed GLD can improve the achievable rate of the SU compared with the traditional SDP method. Besides, the NPSP algorithm was proposed to realize the discretization of phase shifts about the reflecting coefficients at the RIS in practical system.  The RIS can amplify the effective signals and suppress the interference of the SU by empowering the radio environments of wireless channels, which improves the SE performance of the SSS coexisting with multiple PNs significantly.

%


\end{document}